\documentclass{article}
\usepackage{amsfonts,amssymb,amsmath}
\usepackage[dvips]{epsfig}
\usepackage[T1]{fontenc}
\usepackage[latin1]{inputenc}
\usepackage{graphicx}
\usepackage[english]{babel}
\usepackage{amsmath}
\usepackage{amssymb}
\usepackage{amsfonts}
\usepackage{color}

\begin{document}

\title{\bf  Black Hole Solutions Surrounded by Perfect Fluid in Rastall Theory}
\author{Y. Heydarzade$$\thanks{%
email: heydarzade@azaruniv.edu} ~~and ~~F. Darabi$$\thanks{%
email: f.darabi@azaruniv.edu, Corresponding author} ,
\\{\small Department of Physics, Azarbaijan Shahid Madani University, Tabriz, Iran}}
\maketitle
\begin{abstract}
In this work, we obtain uncharged$\setminus$charged  Kiselev-like black holes as a  new class of black hole solutions surrounded by perfect fluid in the context of Rastall theory.   Then, we study the specific cases of the uncharged$\setminus$charged black holes surrounded by regular matter like dust and radiation, or exotic
matter like quintessence, cosmological constant and phantom fields. By comparing
the  Kiselev-like black hole solutions  in Rastall
theory with the Kiselev black hole solutions in GR, we find an effective perfect fluid behaviour for the black hole's surrounding field.
It is shown that the corresponding effective perfect fluid
has interesting characteristic features depending on the different ranges
of the parameters in Rastall theory. For instance, Kiselev-like black holes surrounded by regular matter in Rastall theory may  be considered as
Kiselev black holes surrounded by exotic matter in GR, or Kiselev-like black holes surrounded by exotic matter in Rastall theory may  be considered as
Kiselev black holes surrounded by regular  matter in GR.
\end{abstract}
  
\section{Introduction}
One of the basic elements of Einstein's general theory of relativity (GR) is the so-called covariant conservation of the
energy-momentum tensor which  via the Noether symmetry theorem leads to the conservation of some globally defined physical quantities. These
conserved quantities appear as the integrals of the components of the energy-momentum tensor over appropriate space-like surfaces. These  space-like surfaces  admit at least one of the Killing vectors of the background spacetime as their normal. By this way, the total rest energy/mass of a physical
system is conserved in the context of GR.  On the other hand, some GR based new modified theories  have been proposed that relax the condition of covariant
energy-momentum conservation. 
One of these possible modification of the general theory
of relativity was introduced by P. Rastall in 1972 \cite{rastall1, rastall2}. In this
theory,  the usual conservation law expressed by the null
divergence of the energy-momentum tensor, i.e ${T^{\mu\nu}}_{;\mu}=0$,  is questioned. Then, a non-minimal coupling of matter fields to geometry is considered where
the divergence of $T_{\mu\nu}$
is proportional to the gradient of the
Ricci scalar, i.e ${T^{\mu \nu}}_{;\mu}\propto R^{,\nu}$, such that the usual conservation law is
recovered in the flat spacetime. This can be understood as a direct accomplishment of the Mach principle representing
that the inertia of a mass distribution is dependent on the mass and energy content of the external spacetime \cite{makhjoon}. The main argument in
favor of such a proposal is that the usual conservation
law on  $T_{\mu\nu}$ is tested
only in the flat Minkowski space-time or specifically in
a gravitational weak field limit. Indeed, this theory reproduces
 a phenomenological way for distinguishing features of quantum
effects in gravitational systems, i.e the violation of the classical
conservation laws \cite{Violation, conserv1, conserv2}, which is also reported
in   the $f(R, T)$ \cite{f(rt)**} and $f(R,\mathcal{L}_m)$ \cite{f(t)**} theories, where $R, T$ and $\mathcal{L}_m$ are the Ricci
scalar,  trace of the energy-momentum tensor and  the Lagrangian of the matter
sector, respectively. 
Also,  the  condition ${T^{\mu\nu}}_{;\mu}\neq0$ is 
phenomenologically  confirmed by the particle  creation process  in  cosmology \cite{particle1, particle2, particle3, particle4, particle5, Calogero1, Calogero2,
Velten}. In this regard, 
  the Rastall  theory can 
be  considered as a good candidate for classical  formulation  of  the particle  creation 
through its non-minimal coupling \cite{particle4, string}.  Moreover,  some  astrophysical  analysis  including the evolution of the neutron stars and cosmological data do not 
reject  this  modified theory  \cite{astrophys1, astrophys2, astrophys3}.
Specially, in \cite{astrophys1}  it is shown that the restrictions on the Rastall
geometric parameters are of the order of $\leq1\%$ with respect to the corresponding
value of the Einstein GR. In other words, the results in \cite{astrophys1} confirm that the Rastall theory is a viable theory in the sense that the deviation of
any extended theory of gravity from the standard GR must be weak,  to pass the solar system tests.  Some 
studies on  the various aspects of  this  theory  in  the 
context  of  current  accelerated expansion phase  of  the  universe as well
 other cosmological problems can  be 
found  in \cite{particle4, cosmo1, cosmo2, cosmo3, cosmo4, cosmo5, Hooman0,
al1, al2}. Also, some research works are dedicated to incorporate this theory with the Brans-Dicke and 
scalar-tensor  theories  of  gravity \cite{brans, brans1, scalar}. A  modified Brans-Dicke  theory incorporating Rastall's  assumption, namely a nonzero divergence
of the energy-momentum tensor, is introduced in \cite{smalley1, smalley2} which results in a class of viable theories with consistent field
equations and gauge conditions. The implications of Rastall  assumption in Kaluza-Klein theory and in inflationary cosmologies have been investigated in \cite{KK,
inflation}. It  is also shown that this theory regenerates 
some  loop quantum cosmological features of the universe expansion \cite{loop}.
Apart from the cosmological solutions, any
modified theory must also provide the solutions associated to the stellar and
black hole configurations. In this line, some neutron star, black hole and wormholes solutions in the context of Rastall theory
are obtained in \cite{astrophys1, Yaghoub, Hooman1, Hooman2, oliv, bron}.
{Also, a generalized version  of Rastall theory is recently proposed which shows an agreement with the cosmic
accelerating expansion \cite{new}. In this regard, a dynamical factor for the proportionality
of the energy-momentum tensor divergence and Ricci scalar divergence is considered.  It is shown that this consideration leads to a transition from the matter dominated era to the current accelerating phase of the universe representing
an agreement with some
previous observations \cite{new1, new2, new3}.} Finally, it should be mentioned that
although  Smalley  first tried  to get a  Lagrangian  for   a prototype Rastall theory of gravity,  with a variable gravitational constant \cite{smalley},
  but this theory has been suffered from the lack
of a consistent Lagrangian structure. 
This fact is known as the major  drawback of this theory. But, recently a
Lagrangian formulation for this theory is provided which may motivate the
people to consider this theory more serious than before \cite{lagjoon}. Besides, this theory possesses a rich structure that may be  connected with some fundamental
aspects of a  complete theory of gravity and there are some points in favor
of this theory. First of all, as mentioned before,
the usual energy-momentum conservation law of Einstein's special relativity
(SR) can be
generalized to the  curved spacetime in some different ways, including the
appropriate geometric terms. Indeed, GR theory is one of the possible
extension of SR to the curved spacetime  by simply replacing the standard derivative with a covariant derivative, as the minimal  generalization.   Moreover, the classical form of the energy-momentum tensor must be modified
by introducing quantities related to the curvature of the spacetime when  the quantum effects are taking into account \cite{Violation}.  Also, due to the chirality of the
quantum modes,  the propagation of quantum fields
in the spacetimes possessing horizons may lead to the violation of the classical conservation law which result in the  so-called gravitational anomaly effect
\cite{anomally}.
In this regard, Rastall theory can be a good phenomenological
candidate in order to take into account the effects of quantum fields in curved spacetime  in a  covariant approach.  Although,   there is no action leading to the Rastall equations by implementing the variational principle, but it is possible to obtain such an
action by introducing an external field  in the Einstein-Hilbert action through a Lagrange multiplier. There are other geometrical models such as the well known Weyl
geometry which may result in  the field equations  similar to the  Rastall's
field
equations \cite{smalley, almeidia}. 

On the other hand, the direct local impacts of cosmic backgrounds upon
the known black hole solutions have been paid attention recently. It is shown by Babichev et al \cite{babichev} that for a universe filled by phantom field, the black hole
mass diminishes due to the accreting particles of the phantom
scalar field into the central black hole. But this is a
global impact indeed. The local changes in the spacetime
geometry next to the central black hole can be obtained by a
modified metric including the surrounding space time of the
black hole.
In this regard,
an analytical static spherically symmetric solution to Einstein filed equations has been obtained by Kiselev \cite{Kiselev}. This
solution is characterized by the equation of state parameters of the black hole
surrounding fields which generally can be dust, radiation or a dark energy
component \cite{Kiselev, majeed}. In \cite{majeed}, a Reissner-Nordström black hole surrounded by radiation and dust and a Schwarzschild black hole surrounded by quintessence, as the special cases of the Kiselev general solution,
their phase transitions as well as their thermodynamical properties are investigated.  The dynamics of a neutral and a charged particle around  the Schwarzschild
black surrounded by a quintessence matter have been discussed in  \cite{jamil}.  The rotating Kiselev solution and Kerr-Newman Kiselev
solution have been also obtained in  \cite{Ghosh, Oteev, Toshmatov, Xu}.
Phase transition, quasinormal modes and Hawking radiation of Schwarzschild black hole in the quintessence field are studied in \cite{Tharanath, Zhang, Chen}. Also, One may
refer to \cite{Ghaderi1, Ghaderi2, Wei1, Wei2, Thomas, Pradhan} for more detail in thermodynamical analysis of the Schwarzschild, Reissner-Nordström and Reissner-Nordström-AdS black holes in a quintessence background.

 The essence of the Rastall theory is associated to
the high curvature environments and consequently the    black holes  physics
can
provide an appropriate ground in order to investigate this theory in more
details. Therefore,  in this paper, our aim is
to obtain the   surrounded
  Kiselev-like black hole solutions as a  new  class of
non-vacuum black hole solutions  of this theory. The organization of
the paper is as follows. In section 2, the general analytical static spherical symmetric surrounded black hole solutions in Rastall theory is obtained. Then, in the next five subsections $2.1-2.5$,  the special cases of the surrounded uncharged$\setminus$charged black holes by the dust, radiation, quintessence, cosmological constant and phantom fields are addressed. Finally, in section
3, some concluding remarks are represented.   
 
\section{Surrounded   Black Hole Solutions in Rastall Theory}
In this section, we are looking for the general non-vacuum spherically symmetric static uncharged$\setminus$charged black hole solutions
in the context of the Rastall theory of gravity. Based on the Rastall's hypothesis \cite{rastall1, rastall2}, for a spacetime
with Ricci scalar $R$ filled by an energy-momentum source of $T_{\mu \nu}$, we have
\begin{equation}\label{rastal}
{T^{\mu \nu}}_{;\mu}=\lambda R^{,\nu},
\end{equation}
where $\lambda$ is the Rastall parameter, a measure for deviation from the
standard GR conservation law. Then, the Rastall field  equations can be
written as
\begin{equation}\label{r1}
G_{\mu \nu}+\kappa\lambda g_{\mu \nu}R=\kappa T_{\mu \nu},
\end{equation}
where $\kappa$ is the Rastall
gravitational coupling constant.
This field equations reduce to GR field equations in the limit of $\lambda\rightarrow 0$
and $\kappa=8\pi G_N$ where $G_N$ is the Newton gravitational coupling constant.

In order to obtain black hole solutions, we consider the general spherical symmetric spacetime metric in the standard  
Schwarzschild coordinates as
\begin{equation}\label{metric}
ds^{2}=-f(r)dt^2+\frac{dr^{2}}{f(r)}+r^2d\Omega^2,
\end{equation}
where $f(r)$ is a generic metric function depending on  the radial coordinate $r$ and
$d\Omega^2=d\theta^2+sin^2\theta d\phi^2$ is the two dimensional unit sphere. 
Using this metric, we obtain nonvanishing components of the
Rastall tensor defined as $H_{\mu\nu}=G_{\mu \nu}+\kappa\lambda g_{\mu \nu}R$  as 
\begin{eqnarray}\label{H}
&&{H^{0}}_{0}={G^{0}}_{0}+\kappa\lambda R=-\frac{1}{f}G_{00}+\kappa\lambda R=\frac{1}{r^2}\left(f^{\prime}r-1+f  \right)+\kappa\lambda R,\nonumber\\
&&{H^{1}}_{1}={G^{1}}_{1}+\kappa\lambda R=f G_{11}+\kappa\lambda R=\frac{1}{r^2}\left(f^{\prime}r-1+f  \right)+\kappa\lambda R,\nonumber\\
&&{H^{2}}_{2}={G^{2}}_{2}+\kappa\lambda R=\frac{1}{r^2}G_{22}+\kappa\lambda R=\frac{1}{r^2}\left(rf^{\prime}+\frac{1}{2}r^2 f^{\prime\prime}\right)+\kappa\lambda R,\nonumber\\
&&{H^{3}}_{3}={G^{3}}_{3}+\kappa\lambda R=\frac{1}{r^2 sin^2 \theta}G_{33}+\kappa\lambda R=\frac{1}{r^2}\left(rf^{\prime}+\frac{1}{2}r^2 f^{\prime\prime}\right)+\kappa\lambda R,
\end{eqnarray}
 where the
Ricci scalar reads as
\begin{equation}\label{R}
R=-\frac { 1
}{{r}^{2}}\left({r}^{2}f^{\prime\prime}  +4r f^{\prime}-2+2\,f \right),
\end{equation}
in which the  prime sign represents the derivative with respect to  the radial coordinate $r$.
Then, regarding the nonvanishing components of the Rastall tensor ${H^{\mu}}_{\nu}$, the total energy-momentum tensor supporting this spacetime should have the following diagonal form 
\begin{equation}\label{T}
{T^{\mu}}_{\nu}=
\begin{pmatrix}{T^{0}}_{0} & 0 & 0 & 0 \\
0& {T^{1}}_{1} & 0 & 0 \\
0 & 0 & {T^{2}}_{2} & 0 \\
0 & 0 & 0 & { {T^{3}}_{3}}\\
\end{pmatrix},
\end{equation}
which  must also obey the symmetry properties of the Rastall tensor ${H^{\mu}}_{\nu}$.
 Regarding the equations in (\ref{H}), the equalities  ${H^{0}}_{0}={H^{1}}_{1}$ and ${H^{2}}_{2}={H^{3}}_{3}$ require ${T^{0}}_{0}={T^{1}}_{1}$ and 
 ${T^{2}}_{2}={T^{3}}_{3}$, respectively. 
Then, one can construct  a general total energy-momentum tensor ${{T}^{\mu}}_{\nu}$ possessing
these symmetry properties in the following form
\begin{equation}\label{t**}
{{T}^{\mu}}_{\nu}={E^{\mu}}_{\nu}+{\mathcal{T}^{\mu}}_{\nu},
\end{equation}
where  
  ${E^{\mu}}_{\nu}$ is the trace-free Maxwell tensor given by
\begin{equation}\label{E*}
E_{\mu\nu}={\frac{2}{\kappa}}\left(F_{\mu\alpha}{F_{\nu}}^{\alpha}-
\frac{1}{4}g_{\mu\nu}F^{\alpha\beta}F_{\alpha\beta}\right),
\end{equation}
so that $F_{\mu\nu}$ is the antisymmetric Faraday tensor satisfying
the following vacuum Maxwell equations
\begin{eqnarray}\label{max}
&&{F^{\mu\nu}}_{;\mu}=0,\nonumber\\
&&\partial_{[\sigma} F_{\mu\nu]}=0.
\end{eqnarray}
Considering the spherical symmetry existing in the spacetime metric (\ref{metric})
imposes the only non-vanishing
components of the Faraday tensor $F^{\mu\nu}$ to be $F^{01}=-F^{10}$. Then, from the equations  in (\ref{max}),
one obtains
\begin{equation}\label{f*}
F^{01}=\frac{Q}{r^2},
\end{equation}
where $Q$ is an integration constant playing the role of a electrostatic charge.
Thus, the equations (\ref{metric}), (\ref{E*}) and (\ref{f*}) give the only non-vanishing components of the Maxwell tensor ${E^{\mu}}_{\nu}$ as
\begin{equation}\label{E**}
{E^{\mu}}_{\nu}={\frac{Q^2}{\kappa r^4}}~diag(-1,-1,1,1),
\end{equation}
representing an electrostatic field and clearly possesses the symmetries in ${H^{\mu}}_{\nu}$ tensor.
On the other hand, ${\mathcal{T}^{\mu}}_{\nu}$ describes the energy-momentum
tensor of the surrounding field   defined  as \cite{Kiselev}
 \begin{eqnarray}\label{sur}
&&{\mathcal{T}^{0}}_{0}=-\rho_s(r),\nonumber\\
&&{\mathcal{T}^{i}}_{j}=-\rho_{s}(r)\alpha\left[-(1+3\beta)\frac{r_i r^j}{r_n r^n}+\beta{\delta^{i}}_{j}\right].
\end{eqnarray}
This form of ${\mathcal{T}^{\mu}}_{\nu}$ indicates  that the spatial sector  is proportional to the time sector,  denoting the energy density $\rho_s$,
 with
the arbitrary parameters $\alpha$ and $\beta$ related to the internal structure of the black hole surrounding
field. Here, we used the subscript $``s"$  for denoting the surrounding field which generally can be a dust,
radiation, quintessence, cosmological constant,  phantom field or even any combination
of them.  By taking
the isotropic average over the angles we have \cite{Kiselev} 
\begin{equation}\label{average}
<{\mathcal{T}^{i}}_{j}>=\frac{\alpha}{3}\rho_{s}{\delta^{i}}_{j}=p_{s}{\delta^{i}}_{j},
\end{equation}
since it is  supposed that $<r^{i}r_{j}>=\frac{1}{3}{\delta^{i}}_{j}r_n r^n$. Thus,
one has the barotropic equation of state
for the surrounding field\begin{equation}\label{p}
p_s=\omega_s \rho_s, ~~~\omega_s=\frac{1}{3}\alpha,
\end{equation}
where $p_s$ and $\omega_s$ are the pressure and equation of state parameter,
 respectively. Thus,  the field equations (\ref{H})
with respect to the total energy-momentum tensor in (\ref{t**}), (\ref{E**})
and (\ref{sur}) exactly provide the  principle of additivity and linearity condition supposed in the reference \cite{Kiselev}
which was proposed to determine the  free parameter $\beta$ of the energy momentum-tensor of
the surrounding field as 
\begin{equation}
\beta=-\frac{1+3\omega_s}{6\omega_s}.
\end{equation}
Then, the non-vanishing components of the ${\mathcal{T}^{\mu}}_{\nu}$ tensor can be obtained in the following form
\begin{eqnarray}
&&{\mathcal{T}^{0}}_{0}={\mathcal{T}^{1}}_{1}=-\rho_s,\nonumber\\
&&{\mathcal{T}^{2}}_{2}={\mathcal{T}^{3}}_{3}=\frac{1}{2}(1+3\omega_s)\rho_s,
\end{eqnarray}
which also possess the same symmetries in the Rastall tensor ${H^{\mu}}_{\nu}$. Consequently,
our total constructed energy-momentum tensor in (\ref{t**}) admits all of
the symmetry
properties of ${H^{\mu}}_{\nu}$. One may just consider the ${\mathcal{T}^{\mu}}_{\nu}$
as the only supporting energy-momentum tensor of the Rastall field equations. In this way, the obtained solutions will describe the surrounded uncharged
black hole solutions in the context of the Rastall theory which differ from the ones
in
GR, as we see later.
 Including the Maxwell tensor ${E^{\mu}}_{\nu}$
in ${T^{\mu}}_{\nu}$ provides the possibility of obtaining most general class
of the static 
surrounded charged black hole solutions in the framework of this theory. In the following,
we solve the field equations and obtain its general solution. Then, we address both of the uncharged$\setminus$charged solutions.

The ${H^{0}}_{0}={T^{0}}_{0}$ and ${H^{1}}_{1}={T^{1}}_{1}$ components of the Rastall field equations give
the following differential equation
\begin{equation}\label{e00}
\frac{1}{r^2}\left(rf^{\prime}-1+f  \right)-\frac {\kappa\lambda
}{{r}^{2}}\left({r}^{2}f^{\prime\prime}  +4r f^{\prime}-2+2\,f \right)
=-\kappa\rho_s-\frac{Q^2}{ r^4},
\end{equation} 
and ${H^{2}}_{2}={T^{2}}_{2}$ and ${H^{3}}_{3}={T^{3}}_{3}$ components read as
\begin{equation}\label{e22}
\frac{1}{r^2}\left(rf^{\prime}+\frac{1}{2}r^2 f^{\prime\prime}\right)-\frac {\kappa\lambda
}{{r}^{2}}\left({r}^{2}f^{\prime\prime}  +4r f^{\prime}-2+2\,f \right)
=\frac{1}{2}(1+3\omega_{s} )\kappa\rho_{s}+\frac{Q^2}{ r^4}.
\end{equation}
Thus, we have two unknown functions  $f(r)$ and $\rho_s(r)$ which can
be determined analytically by the above two differential equations.
{Now, by solving the set of differential equations (\ref{e00}) and (\ref{e22})\footnote{{Substituting $\kappa\rho_s(r)$ from differential equation (\ref{e00}) into (\ref{e22}) gives a differential equation for $f(r)$ leading to the solution (\ref{f1}). Then, by substituting the obtained $f(r)$ into the differential equations (\ref{e00}) or (\ref{e22}), one obtains the appropriate form of $\rho_s(r)$ as given by (\ref{rho}) and (\ref{W})}.}, one obtains the following general solution for the metric function}
\begin{equation}\label{f1}
f(r)=1-\frac{2M}{r}+\frac{Q^2}{r^2}
-\frac{N_s}{ r^{\frac{1+3\omega_{s}-6\kappa\lambda(1+\omega_s)}{1-3\kappa\lambda(1+\omega_s)} }},
\end{equation}
with the energy density
in the form of
\begin{equation}\label{rho}
\rho_s (r)=- \frac{3\mathcal{W}_s N_s  }{\kappa r^{\frac{3(1+\omega_s)
-12\kappa\lambda(1+\omega_s)}{1-3\kappa\lambda(1+\omega_s)}}},
\end{equation}
where $M$ and $N_s$ are two integration constants representing the black
hole mass and surrounding field structure parameter, respectively
in which
\begin{equation}\label{W}
{\mathcal{W}_{s}=-\frac{\left(1-4\kappa\lambda\right)
\left(\kappa\lambda\left(1+\omega_s\right)-\omega_s\right)}{\left(1-3\kappa\lambda(1+\omega)
\right)^2},}
\end{equation}
is a geometric constant depending on the Rastall geometric parameters $\kappa$ and $\lambda$
as well as the equation of state parameter $\omega_s$ of the black hole surrounding field. Note that the integration constant $N_s$ represents the characteristic features
of the surrounding field.
{For $\lambda=0$, i.e in the GR limit, we have   $\rho_s (r)=-\frac{3}{\kappa}\mathcal{W}_s N_s  r^{-3(1+\omega_s)}$ where $\mathcal{W}_s= \omega_s$ as in \cite{Kiselev}.
Note that in \cite{Kiselev}, the author used the units of $4\pi G_N=1$ with
a metric possessing a negative signature.}

Regarding the  weak energy condition representing the positivity of  any kind of energy density of the surrounding
field, i.e $\rho_s\geq0$,  imposes the following condition on the geometric
parameters of the theory
\begin{equation}\label{WEC}
\mathcal{W}_s N_{s}\leq0.
\end{equation}
This condition implies that for the surrounding field with geometric parameter
$\mathcal{W}_s>0$, we need $N_{s}<0$ and  conversely for $\mathcal{W}_s<0$, we need $N_{s}>0$. Then, considering that $\mathcal{W}_s$ is given by (\ref{W}), the sign of the
metric parameter $N_s$ depends  on the Rastall geometric parameters $\kappa$, $\lambda$ and the equation of state parameter $\omega_s$ of the surrounding field. In this regard, any set of $\kappa$, $\lambda$ and $\omega_s$ parameters
may admit a different positive or negative $N_s$ values.

Finally, regarding (\ref{f1}), our metric (\ref{metric}) takes the following form 
\begin{equation}\label{METRIC*}
ds^2=-\left(1-\frac{2M}{r}
+\frac{Q^2}{r^2}-\frac{N_s}{r^{\frac{1+3\omega_s -6\kappa\lambda(1+\omega_s)}{1-3\kappa\lambda(1+\omega_s)}}}\right)dt^2
+\frac{dr^2}{1-\frac{2M}{r}+\frac{Q^2}{r^2}
-\frac{N_s}{r^{\frac{1+3\omega_s -6\kappa\lambda(1+\omega_s)}{1-3\kappa\lambda(1+\omega_s)}}}}
+r^2 d\Omega^2.
\end{equation}
In the limit of $\lambda\rightarrow0$ and $\kappa=8\pi G_N$, we recover the Reissner-Nordström black hole surrounded by a surrounding field
in GR which was firstly found by Kiselev \cite{Kiselev} as 
\begin{equation}\label{METRIC2}
ds^2=-\left(1-\frac{2M}{r}+\frac{Q^2}{r^2}-\frac{N_s}{ r^{3\omega_s +1}}  \right)dt^2
+\frac{dr^2}{
1-\frac{2M}{r}+\frac{Q^2}{r^2}
-\frac{N_s}{{r}^{3\omega_s +1}}
}+r^2 d\Omega^2.
\end{equation}Our obtained static solution (\ref{METRIC*}) is  new and possesses some interesting features. By comparing the  metric (\ref{METRIC*}) with the Kiselev metric (\ref{METRIC2})
 in GR, we may obtain an effective equation of state parameter $\omega_{eff}$ for the modification term resulting
from the geometry of the Rastall theory. 

The notion of ``effective equation of state" in Rastall theory has already been studied in
 the cosmological context, where a solution for
the entropy and age problems of the Standard Cosmological Model were provided
\cite{Ref} by considering  Brans-Dicke and Rastall theories of gravity and  performing a perturbative
analysis. It was shown that by introducing an  ``effective equation of state", the Rastall theory
exhibits satisfactory properties at perturbative level in comparison to
the Brans-Dicke theory.

In the next subsections,   the surrounded black hole by the dust, radiation, quintessence, cosmological constant and phantom fields, as the subclasses of the general solution (\ref{METRIC*}), as well
as their
interesting features are studied in detail. At last, we recall that the cases $\kappa\lambda=\frac{1}{4}$
and $\kappa\lambda=\frac{1}{6}$ are generally excluded due to the divergence of the Rastall gravitational coupling constant, as discussed in 
 \cite{rastall1, Hooman2}.

\subsection{The   Black Hole  Surrounded by the Dust Field}
For the dust surrounding field, we set $\omega_d=0$ \cite{Kiselev, Vikman}. Then, the metric (\ref{METRIC*})
takes the following form
\begin{equation}
ds^{2}=-\left(1-\frac{2M}{r}+\frac{Q^2}{r^2}-\frac{N_d}{{r}^{{\frac{1-6\kappa\lambda}{1-3\kappa\lambda}}}}\right)dt^2+ \frac{dr^2}{1-\frac{2M}{r}+\frac{Q^2}{r^2}-\frac{N_d}{{r}^{{\frac{1-6\kappa\lambda}{1-3\kappa\lambda}}}}}+r^2d\Omega^2.
\end{equation}
This  metric  differs from the metric of the surrounded charged black hole by a dust
field    in GR \cite{Kiselev}. One can realize that in GR,  i.e in the limit of $\lambda\rightarrow0 $ and $\kappa=8\pi G_N$, the  black hole in the dust background appears as a charged black hole with an effective
mass $M_{eff}=2M+N_d$. 
Thus, we see that for $\kappa\lambda\neq0$,  the geometric parameters $\kappa$ and $\lambda$ of the Rastall theory can play an important role leading to  distinct solutions relative to GR. Setting $Q=0$ or ${E^{\mu}}_{\nu}$ in
 the total energy-momentum tensor in (\ref{t**}), one arrives at uncharged
 Kiselev-like black hole solutions in the dust background. One can realize
 that for $\kappa\lambda\neq0$ the Rastall correction term never behaves as the mass or charge terms, and introduces a new character to the black
hole, not comparable to
the mass and charge terms.
The presence of such  nontrivial character  can drastically change the thermodynamics,
causal structure
and Penrose diagrams,
due to the Rastall geometric parameters, with respect to those of GR. 

In this case, the geometric parameter $\mathcal{W}_d$ given by the relation
(\ref{W}) reads as 
\begin{equation}
\mathcal{W}_{d}=-\frac{\kappa\lambda\left(1-4\kappa\lambda\right)}{(1-3\kappa\lambda)^2}.
\end{equation}
{Then, regarding the weak energy condition  represented by the relation
(\ref{WEC}),  for $0\leq\kappa\lambda<\frac{1}{4}$ it is required that $N_d>0$,
while for $\kappa\lambda<0~\cup~\kappa\lambda>\frac{1}{4}$, we need $N_d<0$
for the field structure constant. In
this case,  $\mathcal{W}_{d}$ and consequently $\rho_d$ are effectively different from
their GR counterparts such that $\rho_d=
 \frac{3\lambda(1-4\kappa\lambda)N_d}{(1-3\kappa\lambda)^2}
~r^{-\frac{3-12\kappa\lambda}{1-3\kappa\lambda}}$. }

By comparing this metric with the Kiselev metric (\ref{METRIC2})
 in GR, we may obtain an effective equation of state parameter $\omega_{eff}$ for the modification term resulting
from the geometry of the Rastall theory as
\begin{equation}\label{ngn}
\omega_{eff}=\frac{1}{3}\left(-1+\frac{1-6\kappa\lambda}{1-3\kappa\lambda} \right).
\end{equation}
One  may realize that $\omega_{eff}$  can never  be zero (representing a
background
dust
matter) except for the $\kappa\lambda=0$  corresponding to GR limit. Then,
the solutions of this theory are effectively different from those of GR.
Regarding (\ref{ngn}),  two interesting classes are distinguishable as
\begin{itemize}
\item  $\frac{1}{6} <\kappa\lambda<\frac{1}{3}$
which leads to $\omega_{eff}\leq -\frac{1}{3}$. In this case, we have an
effective surrounding fluid with an effective equation of state parameter $\omega_{eff}$,
playing the role of dark energy, which leads to an effective repulsive gravitational effect. Then, regarding this range for $\kappa\lambda$, such  black holes may  contribute to the accelerating
expansion of the universe in the Rastall theory of gravity. 
 In the language of Raychaudhuri equation,
such an effective surrounding fluid violating the strong energy condition  can account for the accelerating expansion of the universe.
Some $\kappa\lambda$ values in the range $\frac{1}{6} <\kappa\lambda<\frac{1}{3}$
and their corresponding effective equation of state $\omega_{eff}$ parameters
accompanied by the
geometric parameters $\mathcal{W}_d$
and $N_d$ are listed
in Table 1.

\begin{table}
\begin{center}
\begin{tabular}{|c|c|c|c|c|}\hline
$\kappa \lambda$ value& $\omega_{eff}$ value&SEC &$\mathcal{W}_d$ value&$N_d$ value\\\hline
$\frac{2}{10}$ & $-\frac{1}{2}$& violated&$-\frac{5}{4}$&  positive \\\hline
$\frac{2}{9}$ & $-\frac{2}{3}$&violated& $-1$& positive\\\hline
$\frac{3}{10}$ & $~-3$& violated&$~~20$ &negative \\\hline
\end{tabular}
\vspace{0.4cm}\\
\caption{Some $\kappa\lambda$ values in the range $\frac{1}{6} <\kappa\lambda<\frac{1}{3}$
and their corresponding effective equation of state $\omega_{eff}$ parameters
accompanied by the
geometric parameters $\mathcal{W}_d$
and $N_d$. }
\end{center}
\end{table}
Interestingly, for $\kappa\lambda=\frac{2}{10}$ and $\frac{2}{9}$, the effective
equation of state $\omega_{eff}$  lies in the quintessence range while for
$\kappa\lambda=\frac{3}{10}$,
it lies in the strong phantom range. This represents the fact that for a
given $\kappa$, the more
large values of $\lambda$, namely the more strong coupling $g_{\mu \nu}R$ in Rastall theory, the more strong acceleration
phase.
\end{itemize}
\begin{itemize}
\item $\kappa\lambda<\frac{1}{6}~ \cup ~ \kappa\lambda>\frac{1}{3}$
which leads to $\omega_{eff}\geq -\frac{1}{3}$. In this case, we have an
effective surrounding fluid with an equation of state parameter respecting
to the strong energy condition possessing the usual attractive gravitational effect.
 This may contribute to the decelerating
expansion or even the contraction of  universe depending on the value of the
effective equation of state parameter
$\omega_{eff}$. 
In the language of Raychaudhuri equation, such a regular effective matter
which respects to the strong energy condition, can justify the deceleration
phase. Some $\kappa\lambda$ values in the range $\kappa\lambda<\frac{1}{6}~ \cup ~ \kappa\lambda>\frac{1}{3}$
and their corresponding effective equation of state $\omega_{eff}$ parameters
accompanied by the
geometric parameters $\mathcal{W}_d$
and $N_d$ are listed
in
Table 2.

\begin{table}
\begin{center}
\begin{tabular}{|c|c|c|c|c|}\hline
$\kappa \lambda$ value & $\omega_{eff}$ value&SEC&${\mathcal{W}_d}$ value
&$N_d$ value \\\hline
$\frac{1}{8}$ & $-\frac{1}{5}$& respected&$-\frac{32}{25}$& positive \\\hline
$\frac{1}{9}$ & $-\frac{1}{6}$&respected &$-\frac{5}{4}$ & positive \\\hline
$\frac{1}{10}$ & $-\frac{1}{7}$&respected &$-\frac{60}{49}$ & positive\\\hline
$\frac{4}{10}$ & $2$&respected &$15$ &negative \\\hline
$\frac{1}{2}$ & $1$&respected &$4$ &negative \\\hline
$1$ &$\frac{1}{2}$ &respected & $\frac{3}{4}$ &negative \\\hline
\end{tabular}
\vspace{0.4cm}\\
\caption{Some $\kappa\lambda$ values in the range $\kappa\lambda<\frac{1}{6}~ \cup ~ \kappa\lambda>\frac{1}{3}$
and their corresponding effective equation of state $\omega_{eff}$ parameters
accompanied by the
geometric parameters $\mathcal{W}_d$
and $N_d$.}
\end{center}
\end{table}
Interestingly, for $\kappa\lambda=\frac{1}{2}$, the effective
equation of state $\omega_{eff}=1$  belongs to the stiff matter possessing
very strong attractive gravitational effect. 
\end{itemize} 
\subsection{The  Black Hole Surrounded by the Radiation Field}
For the radiation surrounding field, we set $\omega_r=\frac{1}{3}$ \cite{Kiselev,
Vikman}. Then, the metric (\ref{METRIC*})
takes the following form
\begin{equation}
ds^{2}=-\left(1-\frac{2M}{r}+\frac{Q^2-N_{r}}{r^2}
\right)dt^2
+\frac{dr^2}{1-\frac{2M}{r}
+\frac{Q^2-N_{r}}{r^2}}+r^2d\Omega^2.
\end{equation}
It is interesting that this case is the same as in GR and the geometric effects of the Rastall
parameters do not appear for a black hole surrounded by the radiation
field \cite{Kiselev}.  {Also, the geometric parameter $\mathcal{W}_r$ given by the relation
(\ref{W}) reads as
}
\begin{equation}\label{radi}
{\mathcal{W}_{r}=\frac{1}{3}},
\end{equation}
and consequently with regard to the weak energy condition for this case, represented by the relation
(\ref{WEC}), { it is required that $N_r<0$ for the radiation field structure
parameter.  Then, by defining the positive structure parameter $\mathcal{N}_r=-N_r$,} we have 
\begin{equation}
ds^{2}=-\left(1-\frac{2M}{r}+\frac{Q^2+\mathcal{N}_{r}}{r^2}\right)dt^2+\frac{dr^2}{1-\frac{2M}{r}
+\frac{Q^2+\mathcal{N}_{r}}{r^2}}+r^2d\Omega^2,
\end{equation}
which is the metric of the well known
 Reissner-Nordström black hole with an effective charge $Q_{eff}=\sqrt{Q^{2}+\mathcal{N}_r}$.
This result is interpreted as the positive contribution of the characteristic feature
of the surrounding radiation field  to the effective charge of the black hole. The appearance of effective charge in the black hole solution cannot
change the causal structure and Penrose diagrams of this black hole solution, in comparison to the Reissner-Nordström black hole.

Setting $Q=0$ or switching off the electrostatic energy-momentum tensor
${E^{\mu}}_{\nu}$ in
 the total energy-momentum tensor in (\ref{t**}), one arrives at the
 Kiselev black hole solutions in the radiation background. In that case,
the resulting metric will be  the Reissner-Nordström black hole with the charge
term $\mathcal{N}_{r}$.    {Also, note that for a radiation background, not only the  metric and the geometric parameter $\mathcal{W}_r$ are the same as in GR  but  also  
the energy density  $\rho_r$ of the background radiation has the same  form
in
comparison to the GR's as $\rho_r = \frac{\mathcal{N}_{r}}{\kappa
r^{4}}$.
It is seen that the value of  radiation energy density $\rho_r$ of the background 
depends not only on the characteristic feature of the surrounding radiation
field $\mathcal{N}_{r}$, but also it depends on the gravitational constant  of the Rastall theory $\kappa$. In general, Rastall's gravitational constant may differs from the
Newton gravitational constant. However, if one sets $\kappa=8\pi G_N$ as in GR, the corresponding energy densities in both of these theories will be
the same. Such a situation occurs also in the cosmological context of the Rastall
theory \cite{cosmo}. In the cosmological setup, the metric solution, i.e the scale factor,
 of the
universe filled by the radiation fluid is exactly the same as in GR. Then, the evolutions
of the universe during the radiation dominated era are the same for both of
the GR and Rastall theories. This fact can be understood by inspecting the original
field equations of the Rastall theory such that for a radiation 
fluid, we have $T = 0$ and $R = 0$ indicating that everything should be the same as
in GR theory.}
\subsection{The  Black Hole Surrounded by the Quintessence Field}
For the quintessence surrounding field, we set $\omega_q=-\frac{2}{3}$ \cite{Kiselev,
Vikman}. Then, the metric (\ref{METRIC*})
takes the following form
\begin{equation}\label{metquint}
ds^{2}=-\left(1-\frac{2M}{r}+\frac{Q^2}{r^2}-
\frac{N_{q}}{r^{\frac{-1-2\kappa\lambda}{1-\kappa\lambda}}}\right)dt^2
+\frac{dr^2}{1-\frac{2M}{r}+\frac{Q^2}{r^2}-\frac{N_{q}}{r^{\frac{-1-2\kappa\lambda}{1-\kappa\lambda}}}}+r^2d\Omega^2.
\end{equation}
This  metric  differs from the metric of the surrounded charged black hole by a quintessence
field in GR \cite{Kiselev}. 
Here,  it is seen that for $\kappa\lambda\neq0$,  the geometric parameters $\kappa$ and $\lambda$ of the Rastall theory can play an important role leading to  distinct solutions, in comparison to GR. In this case, setting $Q=0$ or ${E^{\mu}}_{\nu}=0$ in
 the total energy-momentum tensor in (\ref{t**}), one arrives at uncharged
 Kiselev-like black hole solutions in the quintessence background. Due to
the appearance of nontrivial $N_q$ term, the causal structure and Penrose diagram will be  different from those of Reissner-Nordström black hole in GR.

In this case, the geometric parameter $\mathcal{W}_q$ given by the relation
(\ref{W}) reads as
\begin{equation}\label{WECQ}
{\mathcal{W}_{q}=
-\frac{\left(1-4\kappa\lambda\right)(2+\kappa\lambda)}{3(1-\kappa\lambda)^2}.}
\end{equation}
Then, considering the weak energy condition given by the relation
(\ref{WEC}),{ we require  $N_q>0$
for $0\leq\kappa\lambda<\frac{1}{4}$ 
and $N_q<0$ for $\kappa\lambda>\frac{1}{4}$.} The equation (\ref{WECQ}) shows  that $\mathcal{W}_{q}$ and consequently the corresponding energy density $\rho_q$  effectively differ from
their GR counterparts such that 
{$\rho_q =
\frac{(1-4\kappa\lambda)(2+\kappa\lambda)N_q}{\kappa}~r^{-\frac{1-4\kappa\lambda}{1-\kappa\lambda}} $.}

In this case, by comparing the metric (\ref{metquint}) with the original Kiselev metric (\ref{METRIC2})
in GR,
 one can obtain an effective equation of state parameter $
\omega_{eff}$ for the modification term resulting
by the geometry of the Rastall theory as
\begin{equation}\label{bb}
\omega_{eff}=\frac{1}{3}\left(-1-\frac{1+2\kappa\lambda}{1-\kappa\lambda} \right).
\end{equation}
One  may realize that $\omega_{eff}$  can never be $-\frac{2}{3}$  (the background
quintessence
filed), except for the $\kappa\lambda=0$  which corresponds to the GR limit. Then,
the solutions of this theory are effectively different from GR's. Regarding (\ref{bb}),  two interesting classes are distinguishable as
\begin{itemize}
\item  $-\frac{1}{2} \leqslant\kappa\lambda<1$
which leads to $\omega_{eff}\leq -\frac{1}{3}$. In this case, we have an
effective surrounding fluid with an equation of state parameter  violating the strong energy condition which leads to a repulsive gravitational effect
like as the background quintessence field but with a different repulsive strength. This may contribute to the accelerating
expansion of the universe.  Regarding the appropriate range for $\kappa\lambda$, such  black holes may contribute to the accelerating expansion of the universe in the Rastall theory.
Using the Raychaudhuri equation,
such an effective surrounding quintessence field violating the strong energy condition can justify  the acceleration expansion of the universe.
Some $\kappa\lambda$ values in the range  $-\frac{1}{2} \leqslant\kappa\lambda<1$
and their corresponding effective equations of state $\omega_{eff}$ parameter
with its
behavior, accompanied by the
geometric parameters $\mathcal{W}_q$
and $N_q$ are given
in Table 3.

\begin{table}
\begin{center}
\begin{tabular}{|c|c|c|c|c|}\hline
$\kappa \lambda$ value& $\omega_{eff}$ value& SEC&{$\mathcal{W}_q$} value& $N_q$ value\\\hline
$-\frac{1}{2}$ & $-\frac{1}{3}$& violated& $-\frac{12}{25}$ & positive \\\hline
$\frac{4}{10}$ & $-\frac{4}{3}$& violated& 15 & negative\\\hline
$\frac{1}{2}$ &$-\frac{5}{3}$& violated&  4&negative \\\hline
\end{tabular}
\vspace{0.5cm}\\
\caption{Some $\kappa\lambda$ values in the range  $-\frac{1}{2} \leqslant\kappa\lambda<1$
and their corresponding effective equation of state $\omega_{eff}$ parameters
with their
behaviors, accompanied by the
geometric parameters $\mathcal{W}_q$
and $N_q$.}
\end{center}
\end{table}
Interestingly, the case of $\kappa\lambda=-\frac{1}{2}$ leads to $\omega_{eff}=-\frac{1}{3}$
representing an effective surrounding quintessence field weaker than the one with $\omega_{q}=-\frac{2}{3}$.
In the cosmological setup and through the second Friedmann equation, the
acceleration
equation,  $\omega_{eff}=-\frac{1}{3}$ corresponds to a universe with a uniform expanding velocity, i.e $\ddot a=0$ where $a$ is the scale factor of the
ambient FRW universe filled by an effective field with $\omega_{eff}=-\frac{1}{3}$. For, $\kappa\lambda=\frac{4}{10}$ and $\kappa\lambda=\frac{1}{2}$,
it is seen that
the effective
surrounding field  possesses a repulsive character stronger than the quintessence with $\omega_{eff}=-\frac{4}{3}$ and $\omega_{eff}=-\frac{5}{3}$ which eventually lie in the phantom regime.  
\end{itemize}
\begin{itemize}
\item $\kappa\lambda\leqslant -\frac{1}{2}~ \cup ~ \kappa\lambda>1$
which leads to $\omega_{eff}\geq -\frac{1}{3}$. In this case, we have an
effective surrounding fluid with an equation of state parameter respecting
to the strong energy condition possessing an attractive gravitational effect. This may contribute to the decelerating
expansion or even in contraction of the universe. In this case, although the
black hole is surrounded by the quintessence field with $\omega_q=-\frac{2}{3}$,
however the effective equation of state $\omega_{eff}$ regarding the appropriate range for $\kappa\lambda$
does not belong to the quintessence range. 
For such a regular effective matter
which respects to the strong energy condition, the Raychaudhuri equation
can justify the deceleration
phase or even the contraction of the universe.
Some $\kappa\lambda$ values in the range  $\kappa\lambda\leqslant -\frac{1}{2}~ \cup ~ \kappa\lambda>1$
and their corresponding effective equations of state $\omega_{eff}$ parameters
with their
behaviors, accompanied by the
geometric parameters $\mathcal{W}_q$
and $N_q$ are given
in Table 4.

\begin{table}
\begin{center}
\begin{tabular}{|c|c|c|c|c|}\hline
$\kappa \lambda$ value & $\omega_{eff}$ value& SEC&{$\mathcal{W}_q$} value&$N_q$ value \\\hline
$-1$ & $-\frac{1}{6}$ & respected& $-\frac{5}{16}$& positive \\\hline
$-\frac{3}{2}$ & $-\frac{1}{15}$& respected&$-\frac{28}{121}$ & positive \\\hline
$-2$ & $0$ & respected&$-\frac{9}{49}$  &positive\\\hline
$\frac{3}{2}$ &$\frac{7}{3} $&respected&  $\frac{20}{49}$ &negative \\\hline
$2$ & $\frac{4}{3}$& respected& $\frac{7}{25}$ &negative \\\hline
$\frac{5}{2}$ & $1$& respected& $\frac{36}{169}$ &negative \\\hline
\end{tabular}
\vspace{0.5cm}\\
\caption{Some $\kappa\lambda$ values in the range  $\kappa\lambda\leqslant -\frac{1}{2}~ \cup ~ \kappa\lambda>1$
and their corresponding effective equation of state $\omega_{eff}$ parameters
with their
behaviors, accompanied by the
geometric parameters $\mathcal{W}_q$
and $N_q$.}
\end{center}
\end{table}
In this case, the Rastall's correction term in metric (\ref{metquint}) can
never behave as the charge
term, i.e as $\frac{1}{r^2}$,  to increases or decrease the charge's effect. {But interestingly for $\kappa\lambda=-2$,
which
leads to  the effective
equation of state $\omega_{eff}=0$  representing an effective dust  matter,
it exactly behaves like the mass term,  i.e $\frac{1}{r}$. The sign of metric parameter $N_q $ for $\kappa\lambda=-2$ is positive and consequently, the correction
term contributes to increase  the effect of  Schwarzschild mass term}. {A similar but reverse effect is reported in \cite{babichev} in which  for a universe filled by a phantom field, the black hole
mass smoothly decreases due to the accreting particles of the phantom
scalar field into the central black hole}. This fact can be investigated for
the case of a universe filled by a quintessence filed, which is out of the
scope of the present paper. Also,  $\kappa\lambda=\frac{5}{2}$ leads to the equation of state
parameter $\omega_{eff}=1$ denoting
a stiff matter. 
In conclusion, it is seen that
although the surrounding field is an essentially quintessence  but the effective
 field is not  the quintessence like filed, possessing a negative equation of
state parameter, rather it can behave effectively as dust or even
stiff matter possessing a zero or positive equation of
state parameters, respectively.
\end{itemize}

\subsection{The  Black Hole Surrounded by the Cosmological Constant  Field }
For the cosmological constant surrounding field, we set $\omega_c=-1$ \cite{Kiselev,
Vikman}. Then, the metric (\ref{METRIC*})
takes the following form
\begin{equation}\label{mCC}
ds^{2}=-\left(1-\frac{2M}{r}+\frac{Q^2}{r^2}-N_cr^2\right)du^2+\frac{dr^2}{1-\frac{2M}{r}+\frac{Q^2}{r^2}-N_cr^2}+r^2d\Omega^2.
\end{equation}
It is interesting that this case is the same as  what was already obtained in GR by
Kiselev \cite{Kiselev}. Then, the  Rastall and Einstein theories behave the
same in the cosmological constant background. Here, setting $Q=0$ or switching off ${E^{\mu}}_{\nu}$ in
 the total energy-momentum tensor in (\ref{t**}), one arrives at uncharged
 Kiselev-like black hole solutions in the de Sitter or anti-de Sitter background.

In this case, the geometric parameter $\mathcal{W}_c$ given by the relation
(\ref{W}) reads as

\begin{equation}\label{CC}
{\mathcal{W}_c=-(1-4\kappa\lambda).}
\end{equation}
Then, considering the weak energy condition given by the relation
(\ref{WECQ}), {we require $N_c>0$ for $0\leq\kappa\lambda<\frac{1}{4}$,  
and $N_c<0$ for $\kappa\lambda>1/4$, }corresponding to de Sitter or anti-de Sitter backgrounds, respectively. This shows that the sign of cosmological constant in the Rastall theory depends on its geometric parameters $\kappa$ and $\lambda$. 
Although the form of metric (\ref{mCC})
in this theory  is the same as in GR for cosmological constant background, but the energy density of the cosmological
constant  differs from the GR  due to the geometric features of
the Rastall theory  through the equations (\ref{rho}) and (\ref{CC}). 
In this case, the energy density of the cosmological constant is given {by  $\rho_c=\frac{3(1-4\kappa\lambda)N_c}{\kappa}$.
A similar situation occurs in the cosmological context of the Rastall theory
where the metric solution of the field equations, i.e the scale factor, for the universe dominated by
the cosmological constant has a similar form as in GR, i.e it has an exponential
form. In this case, by comparing the obtained result in \cite{cosmo} as 
$H\propto\sqrt {1-\frac{2}{3}(\frac{3-2\lambda}{2\lambda-1})\rho}$ with the GR's as $H\propto
\sqrt \Lambda$, we see that although the solutions have the same form but the geometric properties of the Rastall theory may affect the energy density of the background cosmological constant.}\footnote{ {One
should note to a little different notation for the field equations in our work and \cite{cosmo}, where the field equations are defined as 
$R_{\mu\nu}-\frac{\lambda}{2}Rg_{\mu\nu}=\kappa T_{\mu\nu}$
and ${T^{\mu\nu}}_{;\mu}
=\frac{1-\lambda}{2\kappa}T^{;\nu}$}.}

\subsection{The  Black Hole Surrounded by the Phantom Field }
For the phantom surrounding field, we set $\omega_p=-\frac{4}{3}$ \cite{Vikman}. Then, the metric (\ref{METRIC*})
takes the following form
\begin{equation}\label{eer}
ds^{2}=-\left(1-\frac{2M}{r}+\frac{Q^2}{r^2}-\frac{N_p}{r^{\frac{-3+2\kappa\lambda}{1+\kappa\lambda}}}\right)dt^2
+\frac{dr^2}{1-\frac{2M}{r}-\frac{N_p}{r^{\frac{-3+2\kappa\lambda}{1+\kappa\lambda}}}}+r^2d\Omega^2.
\end{equation}
This  metric  differs from the metric of the surrounded charged black hole by a phantom
field in GR \cite{Kiselev}. 
For $\kappa\lambda\neq0$,  the geometric parameters $\kappa$ and $\lambda$ of the Rastall theory  plays an important role leading to  distinct solutions in comparison to GR. Also, setting $Q=0$ or switching of ${E^{\mu}}_{\nu}$ in
 the total energy-momentum tensor in (\ref{t**}), one arrives at uncharged
 Kiselev-like black hole solutions in the phantom background. Due to
the appearance of nontrivial $N_p$ term, the causal structure and Penrose diagram will be  different from those of Reissner-Nordström black hole in GR.

In this case, the geometric parameter $\mathcal{W}_p$ given by the relation
(\ref{W}) reads as

\begin{equation}\label{kkl}
{\mathcal{W}_{p}=-\frac{1}{3}\frac{\left(1-4\kappa\lambda\right)(4-\kappa\lambda)}{(1+\kappa\lambda)^2}}.
\end{equation}
Then, considering the weak energy condition given by the relation
(\ref{WEC}), we require $N_p>0$  for $0\leq\kappa\lambda<\frac{1}{4}~\cup ~ \kappa\lambda>4$ and
$N_p<0$ for $\frac{1}{4}<\kappa\lambda<4$. {The equation (\ref{kkl}) shows  that $\mathcal{W}_{p}$ and consequently the corresponding phantom energy density $\rho_p$  effectively differs from
their GR counterparts such that $\rho_p=\frac{\left(1-4\kappa\lambda\right)(4-\kappa\lambda)}{\kappa(1+\kappa\lambda)^2} N_p
~r^{\frac{1-4\kappa\lambda}{1+\kappa\lambda}} $.
}

By comparing this metric with the Kiselev metric (\ref{METRIC2}),
 we may obtain an effective equation of state parameter $
\omega_{eff}$ for the modification term resulting
from the geometry of the Rastall theory as
\begin{equation}
\omega_{eff}=\frac{1}{3}\left(-1-\frac{3-2\kappa\lambda}{1+\kappa\lambda} \right).
\end{equation}
One  may realize that $\omega_{eff}$  never can be $-\frac{4}{3}$  (the background
phantom field), except for the $\kappa\lambda=0$  corresponding to GR limit.
Then,  two interesting classes are distinguishable as
\begin{itemize}
\item  $-1<\kappa\lambda<\frac{3}{2}$
 leading to $\omega_{eff}\leq -\frac{1}{3}$. Then, we have a surrounding fluid with an effective equation of state parameter $\omega_{eff}$
which violates the strong energy condition resulting in  a repulsive gravitational force. Then, in this range of $\kappa\lambda$, these  black holes may contribute to the accelerating
expansion of the universe in the Rastall theory. For such an effective surrounding quintessence field violating the strong energy condition, the Raychaudhuri equation
 can account for the acceleration expansion of the universe. In Table 5, some $\kappa\lambda$ values in the range  $-1<\kappa\lambda<\frac{3}{2}$
and their associated effective equations of state parameters $\omega_{eff}$ parameter
with their
behaviors,
accompanied by the geometric parameters $\mathcal{W}_q$
and $N_q$ are given.
\begin{table}
\begin{center}
\begin{tabular}{|c|c|c|c|c|}\hline
$\kappa \lambda$ value& $\omega_{eff}$ value& SEC&{$\mathcal{W}_p$} value& $N_p$ value\\\hline
$-\frac{1}{2}$ & $-3$& violated & $-\frac{12}{25}$  &positive \\\hline
$\frac{1}{2}$ & $-\frac{7}{9}$& violated & $4$ & negative\\\hline
$1$ &$-\frac{1}{2}$& violated&  $\frac{3}{4}$ &negative \\\hline
\end{tabular}
\vspace{0.5cm}\\
\caption{Some $\kappa\lambda$ values in the range  $-1<\kappa\lambda<\frac{3}{2}$
and their associated effective equation of state parameters $\omega_{eff}$ parameters
with their
behaviors,
accompanied by the geometric parameters $\mathcal{W}_q$
and $N_q$.}
\end{center}
\end{table}
Interestingly, for $\kappa\lambda=-\frac{1}{2}$, we have  $\omega_{eff}=-3$ which has a repulsive character stronger
than the background phantom field with $\omega_p=-\frac{4}{3}$ while for $\kappa\lambda=\frac{1}{2}$ and $\kappa\lambda=1$, we have an effective field with
 a repulsive character weaker
than the background phantom field with $\omega_p=-\frac{4}{3}$ but still lying in the quintessence
range.
\end{itemize}
\begin{itemize}
\item $\kappa\lambda<-1~ \cup ~ \kappa\lambda\geq\frac{3}{2}$
which leads to $\omega_{eff}\geq -\frac{1}{3}$. In this case, we have an
effective surrounding fluid with an equation of state parameter respecting
to the strong energy condition which leads to a attractive gravitational effect.
 This may contribute to the decelerating
expansion or even in contraction of the universe. In this case, although the
black hole is surrounded by the phantom field with $\omega_p=-\frac{4}{3}$,
but the effective equation of state $\omega_{eff}$ regarding the appropriate range of $\kappa\lambda$
does not belong to the phantom range. This effect may cause the contraction
of  universe filled by such a black holes in the Rastall theory of gravity.
For such a regular effective matter
which respects to the strong energy condition, the  Raychaudhuri equation  can justify the deceleration
phase or even the contraction of the universe. Some $\kappa\lambda$ values in the range $\kappa\lambda<-1~ \cup ~ \kappa\lambda\geq\frac{3}{2}$
and their corresponding effective equations of state $\omega_{eff}$ parameter
with their
behaviors, accompanied by the
geometric parameters $\mathcal{W}_p$
and $N_p$ are given
in Table 6.
\begin{table}
\begin{center}
\begin{tabular}{|c|c|c|c|c|}\hline
$\kappa \lambda$ value & $\omega_{eff}$ value& SEC&{$\mathcal{W}_p$} value&$N_p$ value \\\hline
$-\frac{3}{2}$ & $\frac{11}{3}$ & respected& $-\frac{28}{121}$  &positive \\\hline
$-2$ & $2$&respected& $-\frac{9}{49}$  & positive\\\hline
$-\frac{5}{2}$ &$\frac{13}{9} $&respected& $-\frac{44}{289} $ &positive \\\hline
$2$ & $-\frac{2}{9}$& respected& $\frac{7}{25} $ & negative \\\hline
$\frac{5}{2}$ & $-\frac{1}{7}$&respected& $\frac{36}{169} $  &negative \\\hline
$3$ & $-\frac{1}{12}$& respected& $\frac{11}{64} $&negative \\\hline
$4$ & $0$& respected& $\frac{15}{121}$&negative \\\hline
\end{tabular}
\vspace{0.5cm}\\
\caption{Some $\kappa\lambda$ values in the range $\kappa\lambda<-1~ \cup ~ \kappa\lambda\geq\frac{3}{2}$
and their corresponding effective equation of state $\omega_{eff}$ parameters
with their
behaviors, accompanied by the
geometric parameters $\mathcal{W}_p$
and $N_p$.}
\end{center}
\end{table}
{In this case, the Rastall's correction term in metric (\ref{eer}) can never behave like the charge
term, i.e as $\frac{1}{r^2}$,  but interestingly for $\kappa\lambda=4$,
which
leads to  the effective
equation of state $\omega_{eff}=0$  representing an effective dust  matter,
it exactly behaves like the mass term,  i.e $\frac{1}{r}$. For this case,
the sign of  $N_q $ for $\kappa\lambda=4$ is negative and the correction
term contributes to    decreases  the effect of  Schwarzschild mass term.} Such an  effect is reported in \cite{babichev} in which  for a universe filled by a phantom field approaching to the Big Rip, the black hole
mass gradually decreases due to the accreting particles of the phantom
scalar field into the central black hole.
 In conclusion, it is seen that
although the surrounding field is an essentially phantom field but the effective
surrounding field is not the phantom field, rather it can be effectively
a quintessence,
dust or even
stiff matter. 
\end{itemize}

\section{Conclusion}
 We have obtained general uncharged$\setminus$charged  Kiselev-like black hole solutions surrounded
by perfect fluid   in the context of Rastall
theory.  Then, we have investigated
in more detail the specific cases of the  black holes surrounded by dust, radiation, quintessence, cosmological constant and phantom fields.
In each case, the weak energy condition, representing
a positive energy density, is applied to put constraint on the  physical parameters of this modified theory.   By comparing
the new  term in the metric, resulted from the Rastall theory,  with the Kiselev solution in GR, an effective behaviour for the black hole surrounding field is realized.
It is shown that the effective fluid
has different characteristics through its effective equation of state parameter
$\omega_{eff}$ depending on the $\kappa\lambda$ values. In the case of black hole in a dust background with $\omega_d=0$,
for $\frac{1}{6} <\kappa\lambda<\frac{1}{3}$,
we have $\omega_{eff}\leq -\frac{1}{3}$ violating the strong energy (SEC) condition, while for $\kappa\lambda<\frac{1}{6}~ \cup ~ \kappa\lambda>\frac{1}{3}$, we have
$\omega_{eff}\geq -\frac{1}{3}$ respecting to strong energy condition.  
For a black hole in a quintessence background with $\omega_q=-2/3$,  
we have $\omega_{eff}\leq -\frac{1}{3}$ for $-\frac{1}{2} \leqslant\kappa\lambda<1$
violating the strong energy condition, while $\omega_{eff}\geq -\frac{1}{3}$ for $\kappa\lambda\leqslant -\frac{1}{2}~ \cup ~ \kappa\lambda>1$ respecting to strong energy condition.
In the case of a black hole in phantom background with $\omega_p=-4/3$, for
$ -1<\kappa\lambda<\frac{3}{2}$, 
 we
have  $\omega_{eff}\leq -\frac{1}{3}$, while  for $\kappa\lambda<-1~ \cup ~ \kappa\lambda\geq\frac{3}{2}$, we have
 $\omega_{eff}\geq -\frac{1}{3}$.
For such an effective surrounding fluid violating/respecting the strong energy condition, the Raychaudhuri equation
  can account for the accelerating/decelerating expansion of the universe, respectively.
For each of these special classes, some interesting $\kappa\lambda$ values
and their corresponding $\omega_{eff}$  as well as the defined Rastall geometric
parameters $\mathcal{W}_p$
and $N_p$ are given in the tables 1 to 6. For example, for the black hole
in dust background,  for $\kappa\lambda=\frac{2}{10}$ and $\frac{2}{9}$, the effective
equation of state $\omega_{eff}$  lies in the quintessence range while for
$\kappa\lambda=\frac{3}{10}$,
it lies in the strong phantom regime possessing repulsive gravitational effect. For $\kappa\lambda=\frac{1}{2}$, we
have $\omega_{eff}=1$  which belongs to the stiff matter with stronger
gravitational attraction than the background dust. In the case of
a  black hole in a quintessence background with $\omega_{q}=-\frac{2}{3}$,  the case of $\kappa\lambda=-\frac{1}{2}$ leads to $\omega_{eff}=-\frac{1}{3}$
representing an effective surrounding quintessence field weaker than the background.
 For, $\kappa\lambda=\frac{4}{10}$ and $\kappa\lambda=\frac{1}{2}$,
it is seen that
the effective
surrounding field  possesses a repulsive character stronger than the quintessence with $\omega_{eff}=-\frac{4}{3}$ and $\omega_{eff}=-\frac{5}{3}$, respectively,
 which lie in the phantom regime. Also, for $\kappa\lambda=-2$,
we have $\omega_{eff}=0$  representing an effective dust  field while  $\kappa\lambda=\frac{5}{2}$ leads to the equation of state
parameter $\omega_{eff}=1$ denoting
a stiff matter.
In latter cases, it is seen that
although the surrounding field is an essentially quintessence  but the effective
 field is not  the quintessence like filed, possessing a negative equation of
state parameter, rather it can behave effectively as dust or even
stiff matter possessing a zero or positive equation of
state parameters, respectively. Finally, for a black hole in a phantom background
with $\omega_p =-4/3$, for $\kappa\lambda=-\frac{1}{2}$, we have  $\omega_{eff}=-3$ which has a repulsive character stronger
than the background phantom field, while for $\kappa\lambda=\frac{1}{2}$ and $\kappa\lambda=1$, we have  effective fields with
  repulsive character weaker
than the background phantom field, still  lying in the quintessence
range. For $\kappa\lambda=4$,
we
have $\omega_{eff}=0$  representing an effective dust  field. Then, it is seen that
for the latter cases, although the surrounding field is an essentially phantom field but the effective
surrounding field is not the phantom field, rather it can be effectively
a quintessence,
dust or even
stiff matter. It is predicted that the new terms appearing in the Kiselev-like black holes may cause for some drastic changes in their horizons, causal
structures and thermodynamical aspects, in comparison to the Kiselev black holes in GR. 
 Such study is under work by
the authors and will be reported,
elsewhere.


\end{document}